\documentclass[aps,prd,twocolumn,amsmath,showpacs,superscriptaddress,nofootinbib,nopreprintnumbers,tightenlines]{revtex4-1}

\usepackage{verbatim}
\usepackage[T1]{fontenc}
\usepackage[utf8]{inputenc}
\usepackage[american]{babel}
\usepackage{epsfig}
\usepackage{graphicx}

\usepackage{hyperref}

\usepackage{booktabs}
\usepackage{multirow}
\usepackage{dcolumn}
\usepackage{amsmath}
\usepackage{mathtools}
\usepackage{amsfonts}
\usepackage{amssymb}
\usepackage{epstopdf}
\usepackage{bm}
\usepackage{siunitx}
\usepackage{braket}
\usepackage{enumitem}
\usepackage{soul}
\usepackage[table]{xcolor}
\usepackage{color}
\usepackage{transparent}
\usepackage{pifont}

\definecolor{navyblue}{rgb}{0.0, 0.0, 0.5}
\definecolor{royalblue}{rgb}{0.25, 0.41, 0.88}
\definecolor{cadmiumgreen}{rgb}{0.0, 0.42, 0.24}
\definecolor{blue-violet}{rgb}{0.54, 0.17, 0.89}
\definecolor{darkviolet}{rgb}{0.58, 0.0, 0.83}
\definecolor{teal(colorwheel)}{rgb}{1.0, 0.5, 0.0}

\usepackage{hyperref}
\hypersetup{
    colorlinks=true, 
    linkcolor=royalblue, 
    citecolor=magenta}

\newcommand\ee{\end{equation}}
\newcommand\be{\begin{equation}}
\newcommand\eea{\end{eqnarray}}
\newcommand\bea{\begin{eqnarray}}

\newcommand{\lcdm}{\Lambda\mathrm{CDM}}

\usepackage{booktabs}
\usepackage{multirow}
\usepackage{dcolumn}
\usepackage{colortbl}

\definecolor{magenta(process)}{rgb}{1.0, 0.0, 0.56}

\definecolor{darkspringgreen}{rgb}{0.09, 0.45, 0.27}

\definecolor{royalblue(web)}{rgb}{0.25, 0.41, 0.88}
\begin{document}

\title{No evidence for EDE from Planck data in extended scenarios.}

\author{Emanuele Fondi}
\email{emanuelefondi@icc.ub.edu}
\affiliation{ICC, University of Barcelona, IEEC-UB, Mart\'i i Franqu\`es, 1, E-08028 Barcelona, Spain}
\affiliation{Physics Department and INFN, Universit\`a di Roma ``La Sapienza'', Ple Aldo Moro 2, 00185, Rome, Italy}

\author{Alessandro Melchiorri}
\email{alessandro.melchiorri@roma1.infn.it}
\affiliation{Physics Department and INFN, Universit\`a di Roma ``La Sapienza'', Ple Aldo Moro 2, 00185, Rome, Italy}

\author{Luca Pagano}
\email{luca.pagano@unife.it}
\affiliation{Dipartimento di Fisica e Scienze della Terra, Universit\`a degli Studi di Ferrara and INFN -- Sezione di Ferrara, Via Saragat 1, 44122 Ferrara, Italy}
\affiliation{Institut d'Astrophysique Spatiale, CNRS, Univ. Paris-Sud, Universit\'{e} Paris-Saclay, B\^{a}t. 121, 91405 Orsay cedex, France}

\date{\today}


\begin{abstract}
The latest data release from the ACT CMB experiment (in combination with previous WMAP data) shows evidence for an Early Dark Energy component at more than $3$ standard deviations. The same conclusion has been recently shown to hold when temperature data from the Planck experiment limited to intermediate angular scales ($l \le 650$) are included while it vanishes when the full Planck dataset is considered. However, it has been shown that the full Planck dataset exhibits an anomalous lensing component
and a preference for a closed universe at the level of three standard deviation. It is therefore of utmost importance to investigate if these anomalies could anti-correlate with an early dark energy component and hide its presence during the process of parameter extraction. Here we demonstrate that extended parameters choices as curvature, equation of state of dark energy and lensing amplitude $A_L$ have no impact on the Planck constraints on EDE. In practice, EDE does not solve Planck angular spectra anomalies. This indicates that current CMB evidence for an EDE component comes essentially from the ACT-DR4 dataset. 
\end{abstract}

\maketitle

\section{Introduction}

The past twenty years have revolutionized the field of cosmology. The measurements of luminosity distances of Type Ia Supernovae have provided a surprising evidence for an accelerated expansion of our Universe \cite{SupernovaCosmologyProject:1998vns,SupernovaSearchTeam:1998fmf}. The precise measurements of Cosmic Microwave Background anisotropies have spectacularly confirmed the predictions of acoustic oscillations in the primordial plasma as expected in a perturbed Friedmann-Lemaitre expanding universe with inflationary, adiabatic, initial conditions (see e.g. \cite{Boomerang:2000efg,WMAP:2003elm,Planck:2015fie}).
Finally, observations of Baryon Acoustic Oscillation in the galaxy distribution have not only provided a wonderful confirmation of the model of structure formation but also have significantly increased the constraints on its parameter \cite{SDSS:2005xqv,BOSS:2016wmc}. Thanks to these progresses, a cosmological model has emerged, the so-called $\lcdm$ model, that provides an excellent fit to most of current observations (see e.g. \cite{Planck:2018lbu}). 
While the $\lcdm$ model has been clearly one of the most important achievements in modern cosmology, it is based on several simplifications. Current models of structure formation are generally based on three assumptions, i.e. the existence of dark matter, dark energy, and of primordial inflation. The $\lcdm$ model is grounded on three very specific choices for these components: dark matter is assumed as non-interacting and cold, dark energy is a cosmological constant, and inflation is produced by a single, minimally interacting, slow-rolling, scalar field. Because of these simplifications, it makes sense to consider the $\lcdm$ model as an 'effective' theory, or as an approximated version of a more complete theory still to be discovered. As a matter of fact, with the increase of precision in the experimental data, deviations from the $\lcdm$ model are expected (see e.g. \cite{Peebles:2021gou}).

In this respect, the currently most statistically significant discrepancy between observations in the $\lcdm$ model is the so-called Hubble tension between values of the Hubble constant $H_0$ derived directly from luminosity distances of SN-Ia and indirectly from CMB data (see e.g. \cite{Verde:2019ivm, DiValentino:2020zio,DiValentino:2021izs,Schoneberg:2021qvd} for recent reviews). According to the most recent results this discrepancy has now reached the $5$ sigmas limit \cite{Riess:2021jrx}. While systematic effects can clearly still be present (see e.g. \cite{Efstathiou:2020wxn}), the CMB derived value relies on the assumption of the $\lcdm$ model and several models and/or extensions to it have been proposed in the literature that can in principle solve the Hubble tension (\cite{Knox:2019rjx,DiValentino:2021izs,Schoneberg:2021qvd}).
Among them, one of the most promising solutions is given by an Early Dark Energy component \cite{Karwal:2016vyq,Poulin:2018cxd,Smith:2019ihp}. This component does not only solve the discrepancy between Planck and luminosity distance data, but also accommodates Baryon Acoustic Oscillation data (that, combined with Planck, are also in disagreement with low redshift SN-Ia data). When Planck, BAO and SN-Ia data are combined, the best fit with an EDE component is improved by $\Delta \chi^2\sim -15$ with respect to $\lcdm$, showing a clear preference for EDE from this combined dataset \cite{Smith:2019ihp}. The Planck dataset alone, however does not provide any statistically significant indication for EDE \cite{Hill:2020osr}. CMB angular spectra are able to probe an EDE component in the cleanest way since they constrain directly most of the physics around recombination. The lack of any direct indication from Planck for an EDE component, therefore, does not offer much support to this hypothesis. Recent analyses of the ACT-DR4 CMB experiment, however, have shown an indication for an EDE component at the level of $2$ standard deviations \cite{Hill:2021yec}.
Even more recently, it has been shown that the combination of the ACT-DR4 with Planck temperature data at small and intermediate angular scales ($\ell<650$) shows an indication for EDE at more than $3$ standard deviations while there is no evidence for EDE when the full Planck dataset is considered \cite{LaPosta:2021pgm,Smith:2022hwi}.

It is however well known in the literature  that the full Planck angular spectra show also anomalies at the level of three standard deviations (see discussion in \cite{Planck:2018vyg}). In particular, the Planck angular spectra data show an anomalous lensing amplitude, parameterized by the $A_L$ parameter as in \cite{Calabrese:2008rt}, that is larger than expected at the level of three standard deviations. 
It has been shown that this anomaly has several implications. First of all the Planck CMB data prefers a closed universe at more than three standard deviations and it is therefore in tension at a similar statistical significance with BAO data when curvature is considered
\cite{DiValentino:2019qzk,Handley:2019tkm}.
Secondly, the Planck lensing anomaly also affects neutrino mass constraints, anti-correlating with the $\Sigma m_{\nu}$ parameter and leading to very strong (but, {\it de facto} less reliable,) upper limits (see e.g. \cite{Planck:2018vyg,DiValentino:2021imh}).

It is therefore of utmost importance, especially in light of the recent ACT-DR4+Planck results, to investigate if current anomalies in the Planck CMB angular spectra could also anti-correlate with an EDE component. If the anomaly is indeed due to experimental systematic effects, this could mask the presence of EDE in the Planck data and place too strong constraints as in the case of neutrino masses.

This is therefore the goal of our paper: to clarify if the indication of a lensing anomaly or a positive curvature in the Planck data could mask the presence of an EDE component as identified by ACT-DR4.

Our work is structured as follows: in the next section we will briefly review the EDE model (\ref{subsec:EDE}) considered and our analysis method (\ref{subsec:analysis}), and in section \ref{sec:results} we present our results. In Section \ref{sec:conclusions} we finally present our conclusions.

\section{Method}\label{sec:method}

\subsection{Early Dark Energy model}\label{subsec:EDE}

Let us briefly describe here the EDE model used in our analysis.
Modifications of the early expansion history of the Universe affect the CMB power spectra. The most precisely measured CMB feature, as inferred by \emph{Planck} data, is the angular scale of the sound horizon at last scattering $\theta_s=r^*_s/D^*_A$, where
\begin{equation}
    r^*_s = \int _{z_*} ^\infty \frac{{\rm d} z}{H(z)} c_s(z) \;\; , \;\;    D^*_A = \int _0 ^{z_*} \frac{{\rm d} z}{H(z)} .
\end{equation}
While the sound horizon $r^*_s$ depends on the early evolution of $H(z)$ until last scattering, the angular diameter distance to last scattering $D^*_A$ is related to the late expansion history and roughly scales as $H_0^{-1}$.

As suggested in Ref. \cite{Knox:2019rjx}, one of the most promising solutions to the Hubble tension are those which provide a boost of the expansion rate $H(z)$ in the two decades prior to recombination, decreasing the sound horizon. In this way, in order to preserve the well-constrained $\theta_s$, a lower value of $D^*_A$ is then required, determining a higher CMB-inferred value of $H_0$.

Among the different mechanisms proposed in literature to provide this additional pre-recombination contribution to the expansion rate, in this work we focus on a model which introduces a new scalar component in the energy balance of the Universe, its background evolution being described by the Klein-Gordon equation 
\begin{equation}
\ddot{\phi} + 3H\dot{\phi}+V'(\phi)=0 .
\end{equation}
Due to the Hubble friction, the field is frozen at early time and provides a negative pressure, acting as an Early Dark Energy (EDE) component. After this EDE-like phase, the field starts to evolve under a potential of the form
\begin{equation}
V(\phi)=m^2 f^2 \left[1-cos\left(\frac{\phi}{f}\right)\right]^3 ,
\end{equation}
inspired by ultra-light axion fields with mass $m$ and decay constant $f$. Phenomenologically, this is a good choice from a particle physics point of view, since an extremely low mass is required to trigger, as $H\sim m$, an injection of energy within a short window before recombination \cite{Hill:2020osr}.
The order 3 oscillating potential has been shown to match the current data \cite{Smith:2019ihp} and guarantees that the field effectively dilutes faster than radiation, without affecting the late-time expansion of the Universe.

In order to capture the impact of the introduction of the EDE component, it is convenient to recast the particle physics-related quantities $f$ and $m$ to directly associate them to cosmological observables.

More specifically, let us define:
\begin{itemize}
\item an initial field displacement $\theta_i=\frac{\phi_i}{f}$, with $\phi_i$ initial field value;
\item the critical redshift $z_c$ at which the EDE component energy density reaches its maximum value;
\item the maximal fractional contribution of the EDE to the total energy density, $f_{\rm EDE}=({\rho_{\rm EDE}}/{3M_{\rm Pl}^2 H^2})\rvert_{z_c}$
\end{itemize}

The relation between these parameters and the particle physics- related ones is highly nonlinear and is approximately found through an iterative shooting method.

\subsection{Analysis}\label{subsec:analysis}

We consider EDE in extended parameter space. We parametrize the EDE component with the three extra parameters $f_{\rm EDE}$, $z_c$, $\theta_i$, in addition to the standard $6$ parameters of the $\lcdm$ model given by the  baryon $\omega_b$ and cold dark matter densities $\omega_c$, the primordial spectral index $n_S$ and amplitude of scalar $A_S$ inflationary perturbations, the optical depth at reionization $\tau$ and the angular size of the sound horizon at the last scattering surface $\theta_s$.
We extend this EDE scenario by considering (separately) a variation in the curvature of the universe $\Omega_{\rm K}$, a change in the dark energy equation of state $w$ (assumed as a constant), and in the effective lensing parameter $A_L$.

The theoretical predictions for the power spectra are obtained solving the perturbed Klein-Gordon equation for the evolution of perturbations in the EDE field. For this purpose we use a publicly-available \footnote{\url{https://github.com/cmbant/CAMB/releases/tag/1.2.0}} implementation of the EDE model in the \texttt{CAMB} Boltzmann code \cite{Lewis:1999bs} and refer the reader to \cite{Smith:2019ihp} for further details. In order to constrain the parameter space mentioned above, the predictions of the model are tested against the observational data listed below using a modified version, which includes the EDE model parameters, of the \texttt{CosmoMC} code \cite{Lewis:2013hha}, based on the Metropolis-Hastings Markov Chain Monte Carlo algorithm \cite{Metropolis:1953am,Hastings:1970aa}.

In order to perform the analysis, we choose broad uninformative priors on all the model parameters, except for a limit on $f_{\rm EDE} > 0.01$. This choice is guided by the fact that, as already noted in \cite{Smith:2021ear}, a low evidence for EDE might be related to an issue of sampling volume. More specifically, when $f_{\rm EDE} = 0$, any value of the other EDE parameters results in a situation completely equivalent to $\lcdm$. This implies that the Metropolis-Hastings algorithm may be stuck in this preferred $\lcdm$-like region, not sufficiently exploring the whole parameter space.
In our analysis we consider the following data:

\begin{itemize}

\item {\bf CMB}: Temperature and polarization CMB angular power spectra of the Planck legacy release 2018 \texttt{plikTTTEEE+lowl+lowE}~\cite{Planck:2018vyg,Planck:2019nip}. This serves as our 
baseline data set and is 
included in all data combinations.


\item {\bf BAO}: Baryon Acoustic Oscillation measurements 6dFGS~\cite{Beutler:2011hx}, SDSS MGS~\cite{Ross:2014qpa}, and BOSS DR12 “consensus”~\cite{BOSS:2016wmc}, in the same combination used by the Planck collaboration in~\cite{Planck:2018vyg}.

\item {\bf SN}: Luminosity distance data of $1048$ Type Ia Supernovae from the Pantheon catalog~\cite{Pan-STARRS1:2017jku}.


\end{itemize}

\section{Results}\label{sec:results}

\begin{table*}
\begin{center}
\begin{tabular} {l | l l | l l l}
 & \multicolumn{2}{c|}{$\lcdm$ + $\Omega_{\rm K}$} & \multicolumn{3}{c}{EDE + $\Omega_{\rm K}$ }\\
\hline
\hline
 Parameter & \multicolumn{1}{|c}{CMB} & \multicolumn{1}{c|}{CMB+BAO} & \multicolumn{1}{c}{CMB} & \multicolumn{1}{c}{CMB+BAO} & \multicolumn{1}{c}{CMB+SN} \\
\hline
{\boldmath$\Omega_b h^2   $} & $0.0226\pm 0.0002$ & $0.0224\pm0.0002$ & $0.0229\pm0.0002$ & $0.0226\pm0.0002$ & $0.0227\pm0.0002$\\

{\boldmath$\Omega_c h^2   $} & $0.1181\pm 0.0015$ & $0.1197\pm0.0014$ & $0.1219^{+0.0021}_{-0.0037}$ & $0.1240^{+0.0023}_{-0.0039}$ &  $0.1234^{+0.0025}_{-0.0036}$\\

{\boldmath$100\theta_{MC} $} & $1.0411\pm 0.0003$ & $1.0409\pm0.0003$ & $1.0409\pm0.0004$ & $1.0408\pm0.0004$ & $1.0408\pm0.0004$\\

{\boldmath$\tau           $} & $0.049\pm 0.008$ & $0.055\pm0.008$ & $0.049\pm0.008   $ & $0.057\pm0.008   $ & $0.056\pm0.008$\\

{\boldmath${\rm{ln}}(10^{10} A_s)$} & $3.028\pm 0.017            $ & $3.044\pm 0.016$ & $3.034\pm0.017  $ & $3.055\pm0.017$ & $3.051\pm0.018$\\

{\boldmath$n_s            $} & $0.971\pm 0.005 $ & $0.966\pm0.004$ & $0.980^{+0.006}_{-0.008}   $ & $0.975^{+0.006}_{-0.009}   $ & $0.977^{+0.006}_{-0.008}$\\

{\boldmath$\Omega_{\rm K}$} & $-0.044^{+0.018}_{-0.015}$ & $0.0008\pm0.0019$ & $-0.048^{+0.019}_{-0.016}$ & $0.0004\pm0.0020$ & $-0.007\pm0.006$\\

{\boldmath$f_{\rm EDE}        $} & -- & -- & $<0.055$ & $<0.055$ & $<0.060$\\

{\boldmath$z_c            $} & -- & -- & $6900^{+2000}_{-4000}      $ &   $6200^{+2000}_{-4000}      $              & $6020^{+200}_{-3000}$    \\

{\boldmath$\theta_i       $} & -- & -- & $2.40^{+0.64}_{-0.16}$ & $2.50^{+0.58}_{-0.03}$ & $2.57^{+0.49}_{-0.04}$\\
\hline
$H_0                       $ & $54.4^{+4.0}_{-3.3}             $ & $67.8\pm0.7$ & $54.7^{+3.2}_{-4.2}$ & $69.1^{+0.8}_{-1.2}      $ & $66.2^{+2.1}_{-2.5}$\\

$\Omega_m                  $ & $0.484^{+0.068}_{-0.058}          $ & $0.310\pm0.007$ & $0.493\pm0.064   $ & $0.308\pm0.007$ & $0.336\pm0.021$\\

$S_8                       $ & $0.981\pm 0.048            $ & $0.823\pm0.013$ & $0.996^{+0.051}_{-0.045}   $ & $0.834^{+0.014}_{-0.016}$ & $0.862\pm0.025$\\

\hline
\end{tabular}
\end{center}
\caption{Constraints at $68\%$ C.L.: on cosmological parameters from different data combination assuming a $\lcdm$+$\Omega_{\rm K}$ model (left) and a EDE+$\Omega_{\rm K}$ model (right). As we can see, the Planck-only indication for a closed universe is stable under $\lcdm$ or EDE showing that there is no correlation between curvature and EDE.
\label{tab1}}
\end{table*}

\begin{figure*}
\begin{center}
	\includegraphics[width=1\linewidth]{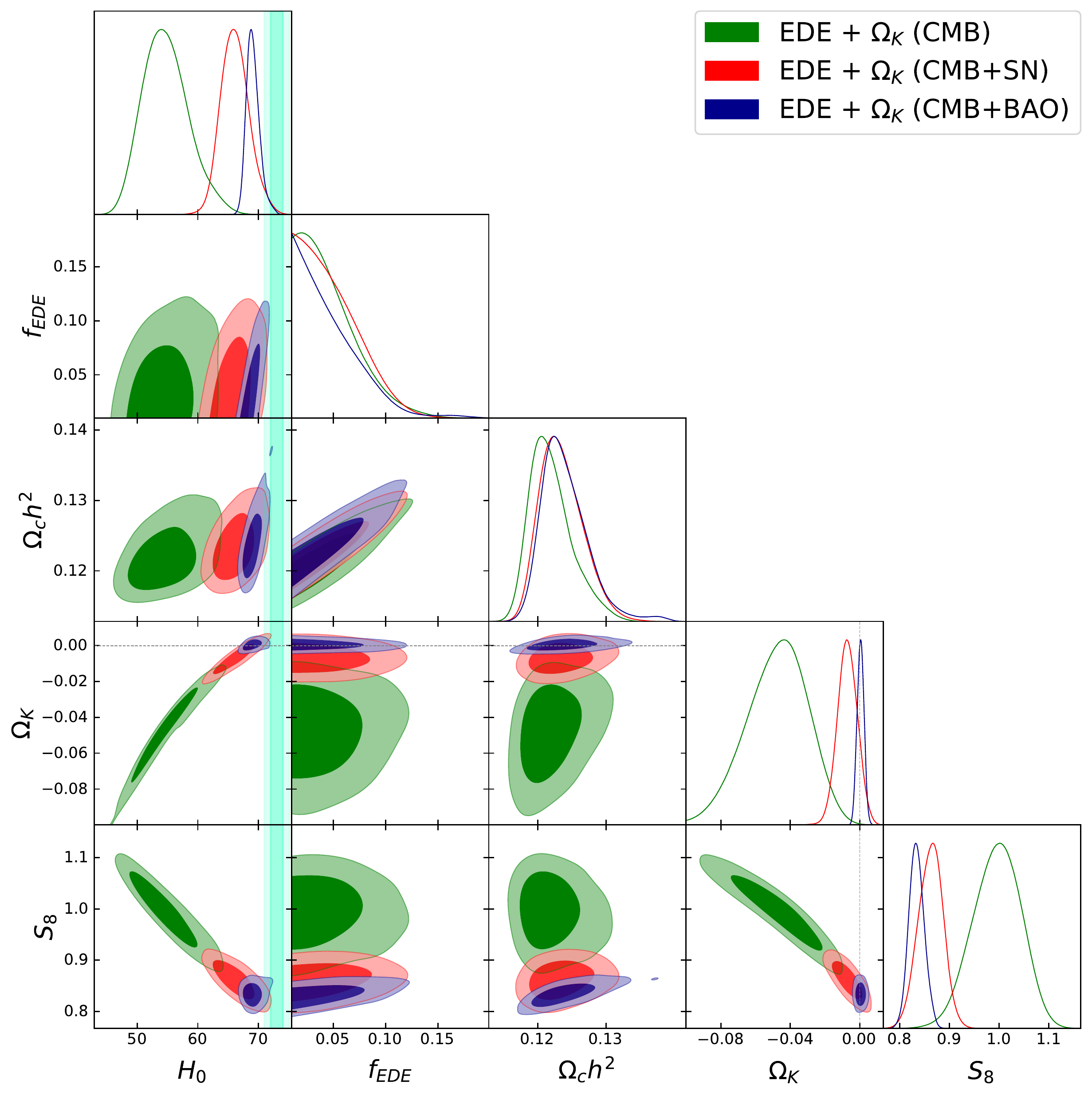}
	\caption{Two-dimensional contour plots at 68\% and 95\% CL for the EDE+$\Omega_{\rm K}$ model. The vertical band shows the S$H_0$ES constraint on $H_0=73.04\pm1.04$ km s$^{-1}$Mpc$^{-1}$ \cite{Riess:2021jrx}. The dashed markers represent the standard value of $\Omega_{\rm K}=0$. As we can see there is a negligible correlation between EDE and $\Omega_{\rm K}$ showing that the Planck indication for a closed universe (and the tension with BAO data) persists under EDE.}
	\label{figure1}
\end{center}
\end{figure*}

Let us first consider a variation in the curvature parameter $\Omega_{\rm K}$ in the framework of the EDE scenario. As already shown in the literature, the Planck angular spectra data prefers a closed universe at the level of about $3$ standard deviations. The same indication is not present in the ACT data \cite{Aiola:2020act}, while this dataset prefers an EDE component also at the level of $3$ standard deviations. The main question we like to answer here is therefore if the curvature anomaly present in the Planck data could affect the Planck constraint on EDE, masking a possible detection. We answer to this question by looking at the parameters constraints reported in Table~\ref{tab1} and in Figure~\ref{figure1}. As we can see there is no significant correlation between the $\Omega_{\rm K}$ and $f_{\rm EDE}$ parameters. The indication for a closed universe from Planck data persists when EDE is considered. The tension between Planck and BAO data (that forces the Planck data to be compatible with a flat universe when combined with it) is also not relieved when EDE is considered. 

\begin{table*}
\begin{tabular} {l | l  l  l | l l l}
 & \multicolumn{3}{c|}{EDE + $w$ } & \multicolumn{3}{c}{EDE + $\Omega_{\rm K}$ + $w$ }\\
\hline
\hline
 Parameter & \multicolumn{1}{c}{CMB} & \multicolumn{1}{c}{CMB+BAO} & \multicolumn{1}{c|}{CMB+SN} & \multicolumn{1}{c}{CMB} & \multicolumn{1}{c}{CMB+BAO} & \multicolumn{1}{c}{CMB+SN}\\ 
\hline
{\boldmath$\Omega_b h^2   $} & $0.0226\pm0.0002$ & $0.0226\pm0.0002$ & $0.0226\pm0.0002$ & $0.0230\pm0.0002$ & $0.0226\pm0.0002$ & $0.0230\pm0.0002$\\

{\boldmath$\Omega_c h^2   $} & $0.1236^{+0.0016}_{-0.0032}$ & $0.1238\pm0.0025$ & $0.1244^{+0.0021}_{-0.0036}$ & $0.1227^{+0.0028}_{-0.0043}$ & $0.1238^{+0.0021}_{-0.0036}$ &  $0.1255^{+0.0027}_{-0.0037}$\\

{\boldmath$100\theta_{MC} $} & $1.0406\pm0.0003$ & $1.0407\pm0.0003$ & $1.0406\pm0.0003$ & $1.0409\pm 0.0004        $ & $1.0407\pm0.0004$ & $1.0409\pm0.0004$\\

{\boldmath$\tau$} & $0.054\pm 0.009          $ & $0.055\pm0.008$ & $0.055\pm0.008   $ & $0.048\pm 0.008          $ & $0.054\pm0.008   $ & $0.049\pm0.008$\\

{\boldmath${\rm{ln}}(10^{10} A_s)$} & $3.050\pm 0.018            $ & $3.054\pm 0.017$ & $3.054\pm0.017  $ & $3.032\pm 0.018            $ & $3.051\pm0.017$ & $3.035\pm0.018$\\

{\boldmath$n_s$} & $0.972^{+0.005}_{-0.006}$ & $0.972\pm0.006$ & $0.971^{+0.005}_{-0.007}   $ & $0.985^{+0.007}_{-0.010}$ & $0.974^{+0.005}_{-0.009}   $ & $0.984^{+0.007}_{-0.009}$\\

{\boldmath$w$} & $-1.47\pm 0.32  $ & $-1.042\pm0.055$ & $-1.033\pm0.038$ & $-0.84^{+0.36}_{-0.18}     $ & $-1.048\pm0.089$ & $-1.27\pm0.10$\\

{\boldmath$\Omega_{\rm K}$} & -- & -- & -- & $-0.067^{+0.012}_{-0.029}  $ & $-0.0008\pm0.0030$ & $-0.036\pm0.010$\\

{\boldmath$f_{\rm EDE}$} & $< 0.037                  $ & $<0.039$ & $<0.042$ & $< 0.074                 $ & $<0.045$ & $<0.065$\\

{\boldmath$z_c$} & $3792^{+390}_{-1400}       $ & $3450^{+500}_{-1000}$ & $3450^{+500}_{-1000}$ & $6728^{+2000}_{-4000}      $ &   $3950^{+400}_{-2000}$& $5460^{+1000}_{-3000}$\\

{\boldmath$\theta_i$} & $2.38^{+0.63}_{-0.16}           $ & $<1.66$ & $<1.73$ & $2.46^{+0.58}_{-0.14}      $ & $<1.92$ & $2.18^{+0.83}_{-1.1}$\\
\hline
$H_0$ & $84^{+10}_{-10}            $ & $69.7\pm1.6$ & $69.4\pm1.2$ & $50.3^{+2.5}_{-6.7}        $ & $69.8\pm1.8$ & $60.6^{+1.9}_{-2.4}$\\

$\Omega_m$ & $0.218^{+0.043}_{-0.070}   $ & $0.303\pm0.012$ & $0.307\pm0.011$ & $0.599^{+0.140}_{-0.096}    $ & $0.303\pm0.014$ & $0.399\pm0.029$\\

$S_8$ & $0.797\pm 0.037            $ & $0.838\pm0.015$ & $0.843\pm0.018$ & $1.034^{+0.047}_{-0.033}   $ & $0.836\pm0.015$ & $0.967\pm0.034$\\
\hline
\end{tabular}
\caption{Constraints at $68\%$ C.L.: on cosmological parameters from different data combination assuming a EDE+$w$ (left) and a EDE+$\Omega_{\rm K}$+$w$ model (right). As we can see, a phantom closed model provides still a better fit to the Planck + SN-Ia data than a flat cosmological constant model. \label{tab2}}
\end{table*}

\begin{figure*}
\begin{center}
	\includegraphics[width=1\linewidth]{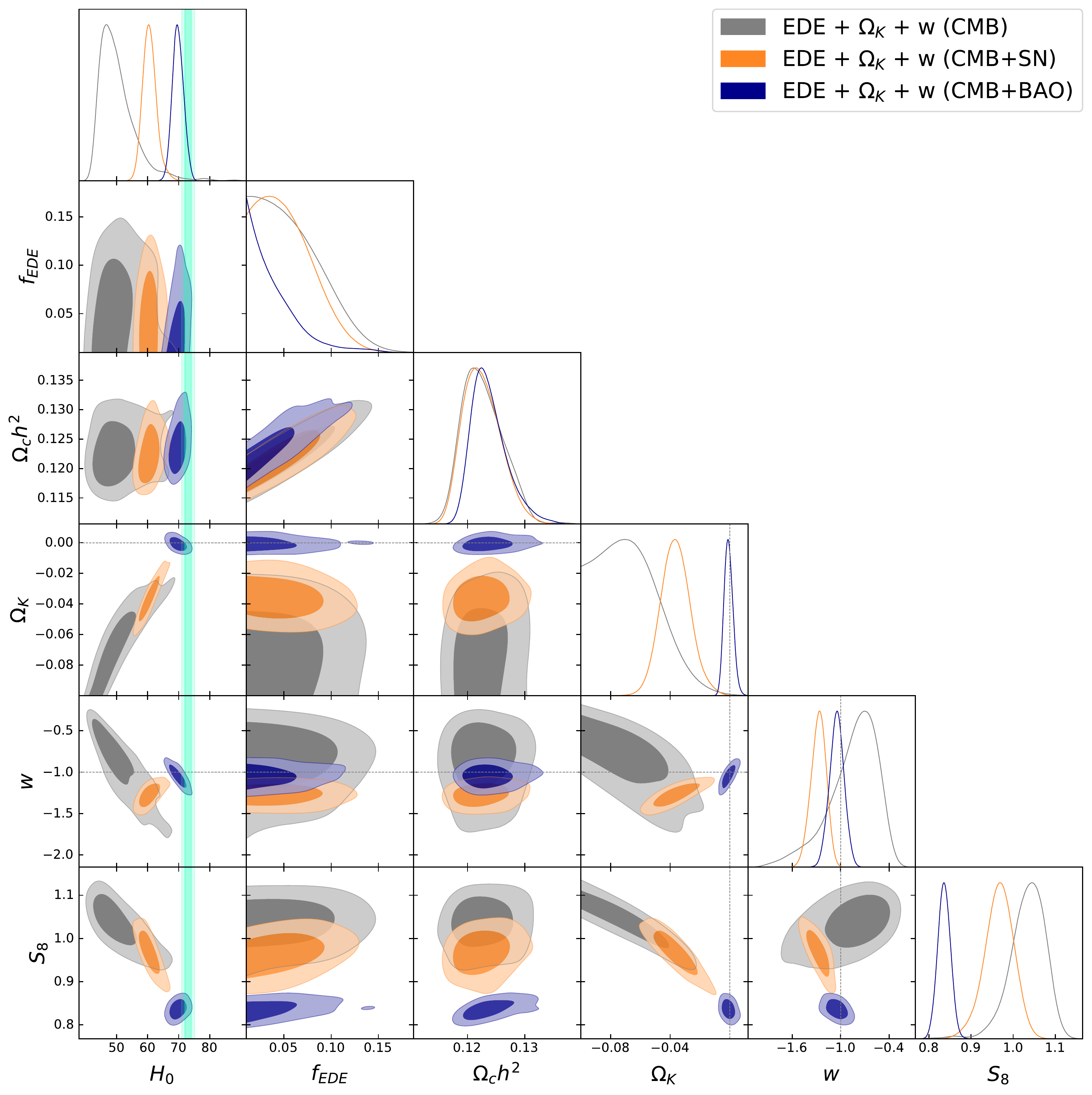}
	\caption{Two-dimensional contour plots at 68\% and 95\% CL for the EDE+$\Omega_{\rm K}$+$w$ model. The vertical band shows the S$H_0$ES constraint on $H_0=73.04\pm1.04$ km s$^{-1}$Mpc$^{-1}$ \cite{Riess:2021jrx}. The dashed markers represent the values of $\Omega_{\rm K}=0$ and $w=-1$. A phantom closed model therefore provides an excellent fit to CMB+SN-Ia data even in the presence of EDE.}
	\label{figure2}
\end{center}
\end{figure*}

As it is well known, the indication for a closed universe from the Planck data is at odds with practically all remaining cosmological observables. This could suggest that either there is an undetected systematic effect in the Planck data, or that additional modifications to the model should be considered. Following this second line of reasoning, it has also been shown that including a variation in the dark energy equation of state, $w$, can make Planck and luminosity distance data in agreement with a closed universe, providing $w<-1$, i.e. also assuming a dark energy phantom model.
While a phantom closed model could appear as an extreme modification to the flat $\lcdm$ model, it has been shown that a Planck+Pantheon analysis prefers this model over $\lcdm$ at the level of three standard deviations.

It is therefore interesting to investigate the bounds on the EDE component also in this extreme scenario. In Table~\ref{tab2} and in Figure~\ref{figure2} we report the constraints obtained in the case of an EDE+$w$ and EDE+$\Omega_{\rm K}$+$w$ scenarios. As we can see, there is no correlation between EDE and the dark energy equation of state. For example, the constraints on $w$ from Planck+BAO are essentially the same in the $\lcdm$ and EDE cases. 

When moving to the EDE+$\Omega_{\rm K}$+$w$ extension we
note that also in this case the constraints on 
$f_{\rm EDE}$ are practically the same as those reported in Table~\ref{tab1} with $w=-1$. 
We could also see that a closed phantom model is preferred from the Planck+SN dataset at about three standard deviations while Planck+BAO data prefer a flat cosmological constant model. As we can better see in Figure~\ref{figure2}, there is indeed a significant tension between the Planck+BAO and Planck+SN results with the corresponding confidence level contours that are completely disjointed in the $\Omega_{\rm K}$-$w$ plane and prefer two significantly different scenarios. Since this tension is present also in the $\lcdm$ model, we can therefore conclude that the introduction of EDE does not help in alleviating the tension between these two datasets in case of a $\Omega_{\rm K}+w$ scenario.

\begin{table*}
\begin{tabular} {l | l l l | l l l}
& \multicolumn{3}{c|}{EDE + $A_L$ } & \multicolumn{3}{c}{EDE + $\Omega_{\rm K}$ + $A_L$ }\\
\hline
\hline
Parameter & \multicolumn{1}{c}{CMB} &  \multicolumn{1}{c}{CMB+BAO} & \multicolumn{1}{c|}{CMB+SN}& \multicolumn{1}{c}{CMB} &  \multicolumn{1}{c}{CMB+BAO} & \multicolumn{1}{c}{CMB+SN}\\
\hline

{\boldmath$\Omega_b h^2   $} & $0.0229\pm 0.0002$ & $0.0229\pm0.0002$ & $0.0229\pm 0.0002$& $0.0229\pm 0.0002$ & $0.0229\pm0.0002$ & $0.0230\pm0.0003$\\

{\boldmath$\Omega_c h^2   $} & $0.1222^{+0.0023}_{-0.0037}$ & $0.1224^{+0.0019}_{-0.0035}$ & $0.1223^{+0.0024}_{-0.0038}$& $0.1225^{+0.0021}_{-0.0042}$ & $0.1220^{+0.0024}_{-0.0036}$ & $0.1224^{+0.0031}_{-0.0038}$\\

{\boldmath$100\theta_{MC} $} & $1.0409\pm 0.0004$ & $1.0409\pm0.0004$ & $1.0408\pm 0.0003$& $1.0409\pm 0.0004$ & $1.0409\pm0.0004$ & $1.0409\pm 0.0003        $\\

{\boldmath$\tau           $} & $0.050\pm 0.008$ & $0.050\pm0.008$ & $0.050\pm 0.008$& $0.050\pm 0.008$ & $0.050\pm0.008$ & $0.051\pm 0.009          $\\

{\boldmath${\rm{ln}}(10^{10} A_s)$} & $3.036\pm 0.018            $ & $3.037\pm 0.018$ & $3.038\pm 0.019            $& $3.037\pm 0.018            $ & $3.035\pm 0.019$ & $3.037\pm 0.019            $\\

{\boldmath$n_s            $} & $0.982^{+0.007}_{-0.009}   $ & $0.979^{+0.006}_{-0.008}   $ & $0.982^{+0.006}_{-0.009}   $& $0.980^{+0.007}_{-0.009}   $ & $0.981^{+0.007}_{-0.009}   $ & $0.986^{+0.008}_{-0.013}$\\

{\boldmath$\Omega_{\rm K}$} & -- & -- &--& $-0.041\pm 0.032        $ & $-0.0007\pm0.0020$ & $-0.004^{+0.006}_{-0.005}$\\

{\boldmath$A_L$} & $1.209\pm 0.070        $ & $1.188\pm0.062$ & $1.205\pm0.071        $& $1.036^{+0.088}_{-0.160}        $ & $1.192\pm0.065$ & $1.192^{+0.061}_{-0.069}   $\\

{\boldmath$f_{\rm EDE}        $} & $<0.064$ & $<0.057$ & $<0.054$& $<0.062$ & $<0.061$ & $< 0.070  $\\

{\boldmath$z_c            $} & $6671^{+2000}_{-4000}      $ &   $6813^{+2000}_{-4000}      $  &  $4449^{+190}_{-1900}      $ &  $6534^{+2000}_{-4000}      $ &   $6752^{+2000}_{-4000}      $   & $5011^{+2000}_{-2000}   $   \\

{\boldmath$\theta_i       $} & $2.49^{+0.54}_{-0.09}     $ & $2.46^{+0.62}_{-0.09}      $ & $2.42^{+0.65}_{-0.14}     $& $2.47^{+0.56}_{-0.09}     $ & $2.46^{+0.57}_{-0.11}      $ & $2.39^{+0.64}_{-0.16}      $\\
\hline
$H_0                       $ & $70.0^{+1.1}_{-1.4}             $ & $69.5^{+0.8}_{-1.1}        $ & $69.9^{+0.9}_{-1.3}             $& $57.9^{+4.7}_{-11}             $ & $69.6^{+0.9}_{-1.2}        $ & $68.5\pm 2.4     $\\

$\Omega_m                  $ & $0.297\pm 0.010$ & $0.302\pm0.006$  & $0.298\pm0.009  $& $0.46\pm 0.13          $ & $0.301\pm0.007$ & $0.312\pm0.021  $\\

$S_8                       $ & $0.805\pm0.020   $ & $0.812\pm0.016   $ & $0.809\pm0.020   $& $0.96^{+0.15}_{-0.09}   $ & $0.809\pm0.016   $ & $0.823\pm 0.029 $\\
\hline
\end{tabular}
\caption{Constraints at $68\%$ C.L.: on cosmological parameters from different data combination assuming a EDE+$A_L$ (left) and a EDE+$\Omega_{\rm K}$+$A_L$ model (right).
\label{tab3}}
\end{table*}

\begin{figure*}
\begin{center}
	\includegraphics[width=1\linewidth]{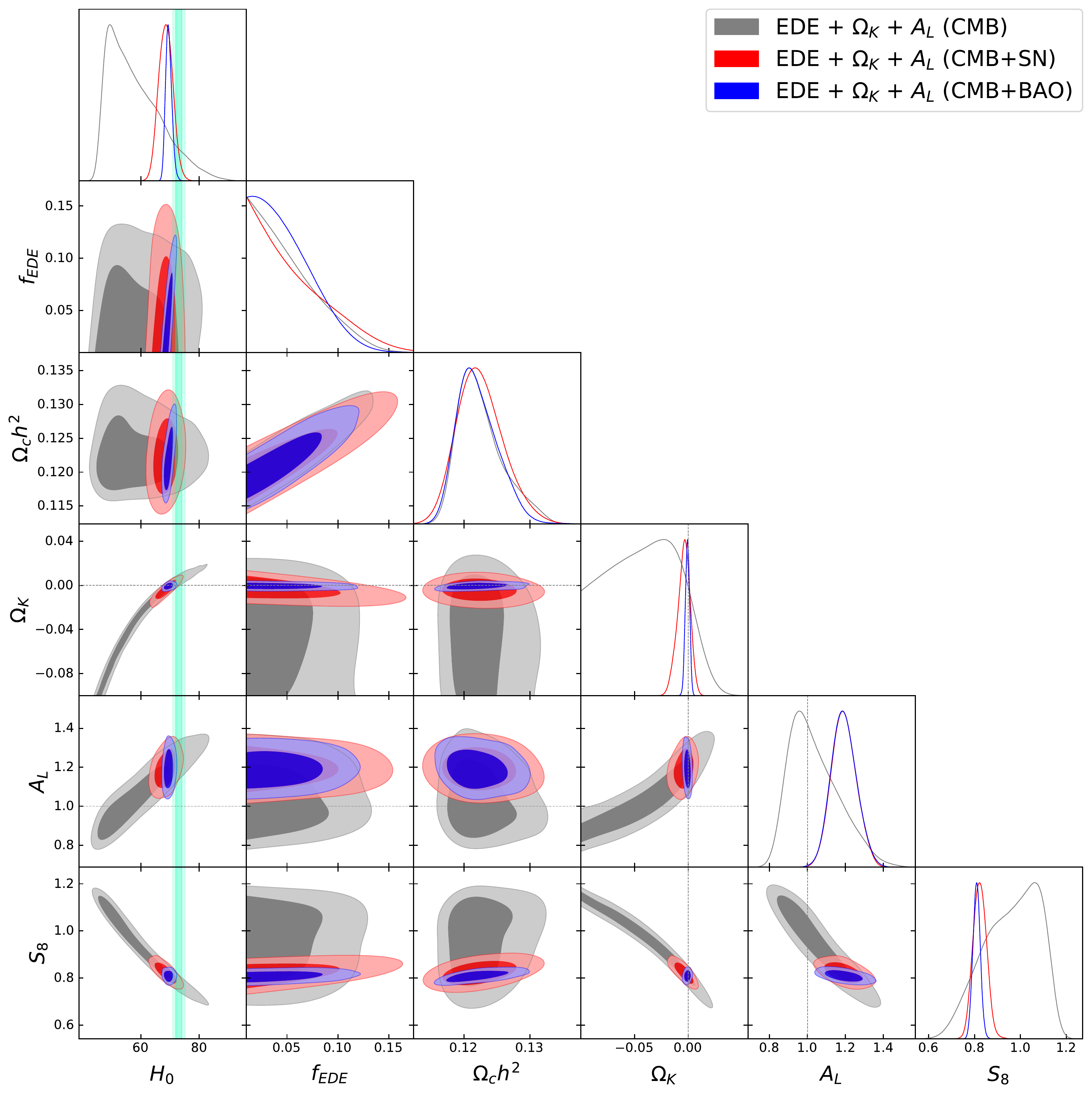}
	\caption{Two-dimensional contour plots at 68\% and 95\% CL for the EDE+$\Omega_{\rm K}$+$A_L$ model. The vertical band shows the S$H_0$ES constraint on $H_0=73.04\pm1.04$ km s$^{-1}$Mpc$^{-1}$ \cite{Riess:2021jrx}. The dashed markers represent the values of $\Omega_{\rm K}=0$ and $A_L=1$. As we can see there is no correlation between EDE and, $A_L$ and $\Omega_{\rm K}$ parameters. The inclusion of the BAO dataset significantly prefers $A_L>1$ solution to the Planck anomalies, leaving the constraints on $f_{\rm EDE}$ as practically unchanged.}
	\label{figure3}
\end{center}
\end{figure*}

Finally, we also study the constraints in the EDE+$\Omega_{\rm K}$+$A_L$ scenario, therefore considering at the same time both changes in the curvature and in the $A_L$ lensing parameter.
We report the constraints obtained in this parameter extension in Table~\ref{tab3} and in Figure~\ref{figure3}.

As already pointed out in the literature, there is a strong degeneracy between curvature and the $A_L$ parameter when only the Planck data is considered. The degeneracy is clearly noticeable by looking at the constraints in the $A_L$ vs $\Omega_{\rm K}$ plane in Figure~\ref{figure3}. The indication for a closed universe from Planck angular spectrum data is clearly driven by the $A_L$ anomaly and this is confirmed also when an EDE component is considered.

However, despite varying both the curvature and the $A_L$ parameter and having therefore a significant variation in the constraints on cosmological parameters with respect to the $\lcdm$ case, the upper limits on the $f_{\rm EDE}$ component are essentially left unchanged. As we can see from Figure~\ref{figure3}, the posteriors on $f_{\rm EDE}$ do not show any indication for a value different from zero.

\section{Conclusions}\label{sec:conclusions}

In this brief paper we have considered constraints on an early dark energy component from the full Planck dataset in an extended parameter space. Since the Planck angular power spectrum data is showing some anomalous value of the lensing amplitude or curvature parameter, we checked if these anomalies could also hide a possible indication for an early dark energy component by anti correlating with the $f_{\rm EDE}$ parameter.
The answer to this question is negative: the Planck constraints on the $f_{\rm EDE}$ parameter are essentially unaffected by variation in lensing, curvature, and dark energy equation of state. Our results therefore suggest that the main indication for a EDE component mostly relies on the ACT-DR4 dataset.


\begin{acknowledgments}
E.F.~acknowledges the support from ”la Caixa” Foundation (ID 100010434, code LCF/BQ/DI21/11860061) and the “Center of Excellence Maria de Maeztu 2020-2023” award to the ICCUB (CEX2019-000918-M funded by MCIN/AEI/10.13039/501100011033).
A.M.~thanks TASP INFN initiative for support. 
L.P.~acknowledges the financial support from the INFN InDark initiative and from the COSMOS network (www.cosmosnet.it) through the ASI (Italian Space Agency) Grants 2016-24-H.0 and 2016-24-H.1-2018.
We acknowledge the use of \texttt{getdist} \citep{Lewis:2019xzd} package, and the use of computing facilities at CINECA.
\end{acknowledgments}

\bibliography{bibfiles.bib}


\vfill
\end{document}